\documentstyle[12pt]{article}
\catcode`\@=11

\def\eqnarray{\let\@currentlabel=\theequation\refstepcounter{equation}
    \global\@eqnswtrue
    \global\@eqcnt\z@\tabskip\@centering\let\\=\@eqncr
    $$\halign to \displaywidth\bgroup\@eqnsel\hskip\@centering
      $\displaystyle\tabskip\z@{##}$&\global\@eqcnt\@ne
       \hfil${{}##{}}$\hfil
      &\global\@eqcnt\tw@ $\displaystyle\tabskip\z@{##}$\hfil
       \tabskip\@centering&\llap{##}\tabskip\z@\cr}
\def\lefteqn#1{\hbox to 4\arraycolsep{$\displaystyle #1$\hss}}

\long\def\@makefntext#1{\parindent 0cm\noindent
\hbox to 1em{\hss$^{\@thefnmark}$}#1}
\def\ra{\rightarrow}
\def\IR{{\hbox{{\rm I}\kern-.2em\hbox{\rm R}}}}
\def\IH{{\hbox{{\rm I}\kern-.2em\hbox{\rm H}}}}
\def\IC{{\ \hbox{{\rm I}\kern-.6em\hbox{\bf C}}}}
\font\cmss=cmss12 \font\cmsss=cmss12 at 12truept
\def\IZ{\relax\ifmmode\mathchoice
{\hbox{\cmss Z\kern-.4em Z}}{\hbox{\cmss Z\kern-.4em Z}}
{\lower.9pt\hbox{\cmsss Z\kern-.36em Z}}
{\lower1.2pt\hbox{\cmsss Z\kern-.36em Z}}
\else{\cmss Z\kern-.4em Z}\fi}
\def\rref#1{(\ref{#1})}
\def\Tr{\hbox{Tr}}

\def\CM{{\cal M}}
\newcommand{\J}{\langle J^z\rangle_\chi}
\newcommand{\beq}{\begin{equation}}
\newcommand{\eeq}{\end{equation}}
\newcommand{\NPB}[1]{{\sl Nucl.~Phys.}~{\bf B#1}}
\newcommand{\Ann}[1]{{\sl Ann.~Phys.}~{\bf #1}}
\newcommand{\CMP}[1]{{\sl Commun.~Math.~Phys.}~{\bf #1}}
\newcommand{\PLB}[1]{{\sl Phys.~Lett.}~{\bf B#1}}

\newcommand{\PRL}[1]{{\sl Phys.~Rev.~Lett.}~{\bf #1}}

\newcommand{\MPLA}[1]{{\sl Mod.~Phys.~Lett.}~{\bf A#1}}
\newcommand{\IJMPA}[1]{{\sl Int.~J.~Mod.~Phys.}~{\bf A#1}}
\newcommand{\IJMPB}[1]{{\sl Int.~J.~Mod.~Phys.}~{\bf B#1}}

\newcommand{\PRD}[1]{{\sl Phys.~Rev.}~{\bf D#1}}

\begin{document}
%
%
%
%
\def\citen#1{%
\edef\@tempa{\@ignspaftercomma,#1, \@end, }
\edef\@tempa{\expandafter\@ignendcommas\@tempa\@end}%
\if@filesw \immediate \write \@auxout {\string \citation {\@tempa}}\fi
\@tempcntb\m@ne \let\@h@ld\relax \let\@citea\@empty
\@for \@citeb:=\@tempa\do {\@cmpresscites}%
\@h@ld}
%
\def\@ignspaftercomma#1, {\ifx\@end#1\@empty\else
   #1,\expandafter\@ignspaftercomma\fi}
\def\@ignendcommas,#1,\@end{#1}
%
%
\def\@cmpresscites{%
 \expandafter\let \expandafter\@B@citeB \csname b@\@citeb \endcsname
 \ifx\@B@citeB\relax 
    \@h@ld\@citea\@tempcntb\m@ne{\bf ?}%
    \@warning {Citation `\@citeb ' on page \thepage \space undefined}%
 \else
    \@tempcnta\@tempcntb \advance\@tempcnta\@ne
    \setbox\z@\hbox\bgroup 
    \ifnum\z@<0\@B@citeB \relax
       \egroup \@tempcntb\@B@citeB \relax
       \else \egroup \@tempcntb\m@ne \fi
    \ifnum\@tempcnta=\@tempcntb 
       \ifx\@h@ld\relax 
          \edef \@h@ld{\@citea\@B@citeB}%
       \else 
          \edef\@h@ld{\hbox{--}\penalty\@highpenalty \@B@citeB}%
       \fi
    \else   
       \@h@ld \@citea \@B@citeB \let\@h@ld\relax
 \fi\fi%
 \let\@citea\@citepunct
}
%
\def\@citepunct{,\penalty\@highpenalty\hskip.13em plus.1em minus.1em}%
%
%
\def\@citex[#1]#2{\@cite{\citen{#2}}{#1}}%
%
%
\def\@cite#1#2{\leavevmode\unskip
  \ifnum\lastpenalty=\z@ \penalty\@highpenalty \fi 
  \ [{\multiply\@highpenalty 3 #1
      \if@tempswa,\penalty\@highpenalty\ #2\fi 
    }]\spacefactor\@m}
\let\nocitecount\relax  
%
\begin{titlepage}
\vspace{.3in}
\begin{flushright}
UCD-93-7\\
DFTUZ 93.5\\
April 1993\\
hep-th/9304081\\
\end{flushright}
\vspace{.2in}
\begin{center}
{\Large\bf
Chern-Simons States\\[1.2ex] and Topologically Massive Gauge Theories}\\
\vspace{.4in}
{
M.~A{\sc sorey},\footnote{\it email: asorey@saturno.cie.unizar.es}\
F.~F{\sc alceto}\footnote{\it email: falceto@cc.unizar.es}\\
       {\small\it Departamento de F\'{\i}sica Te\'orica}\\
       {\small\it Facultad de Ciencias}\\
       {\small\it Universidad de Zaragoza}\\
	{\small\it 50009 Zaragoza}\\
       {\small\it Spain}\\
\vspace{1ex}
{\small and}\\
\vspace{1ex}
S.~C{\sc arlip}\footnote{\it email: carlip@dirac.ucdavis.edu}\\
       {\small\it Department of Physics}\\
       {\small\it University of California}\\
       {\small\it Davis, CA 95616}\\{\small\it USA}}
\end{center}

\vspace{2ex}
\begin{center}
\begin{minipage}[t]{5in}
\begin{center}
{\large\bf Abstract}
\end{center}
{\small
In an abelian topologically massive gauge theory, any eigenstate of
the Hamiltonian can be decomposed into a factor describing massive
propagating gauge bosons and a Chern-Simons wave function describing
a set of nonpropagating ``topological'' excitations.  The energy
depends only on the propagating modes, and energy eigenstates thus
occur with a degeneracy that can be parametrized by the Hilbert
space of the pure Chern-Simons theory.  We show that for a
{\em nonabelian} topologically massive gauge theory, this degeneracy
is lifted: although the Gauss law constraint can be solved with
a similar factorization, the Hamiltonian couples the propagating and
nonpropagating (topological) modes. }

\end{minipage}
\end{center}
\end{titlepage}
\addtocounter{footnote}{-3}

Three-dimensional topologically massive gauge theories \cite{Schon,DJT}
have recently attracted considerable attention as useful models for
condensed matter physics \cite{Wen,Wen2,Zhang,Ho} and string theory
\cite{Kogan,CarKog}.  With actions that combine an ordinary Yang-Mills
kinetic term and a Chern-Simons term, these models exhibit a hybrid
behavior: they contain massive propagating ``photons'' or ``gluons,''
but also display such ``topological'' features as statistical
transmutation.

The propagating sector of a topologically massive gauge theory is
relatively easy to understand in the language of ordinary field
theory.  The behavior of the ``topological'' sector is more
subtle.  For an {\em abelian} theory, it is known that energy
eigenstates are degenerate, with a set of ground states that are
in precise one-to-one correspondence with the states of the associated
pure Chern-Simons theory \cite{Wen,Dunne,Kogan2}.  These degenerate
states determine the ``topological'' features of the theory: in the
spectral decomposition of the propagator, for instance, they account
for the long range Aharonov-Bohm interaction that leads to fractional
spin and statistics \cite{Kogan3}.  (In absence of matter fields,
the degeneracy associated with the topological modes disappears in the
infinite area limit, however \cite{kar}.)

It is an interesting open question whether the correspondence between
degenerate energy eigenstates and states of a Chern-Simons theory
continues to hold in the nonabelian case.  The purpose of this Letter
is to demonstrate  that for nonabelian theories a simple
splitting  of wave functions into ``propagating'' and
``topological'' factors is  no longer possible, and  that
the self-interaction of the gauge field lifts this degeneracy.

Let us consider a topologically massive gauge theory with gauge group
$G$ on a three-manifold $M=\IR\times\Sigma$, where $\Sigma$ is an
arbitrary Riemann surface.  In terms of a Lie algebra-valued gauge
potential $A_\mu$, our action is
\beq
S =
 \int\!d^3x\,\Tr\,\left\{ -{1\over 4\gamma}\sqrt{|g|}F_{\mu\nu}F^{\mu\nu}
      + {k\over 8\pi}\epsilon^{\mu\nu\rho}
      \left(A_\mu(\partial_\nu A_\rho - \partial_\rho A_\nu)
      + {2\over3}A_\mu[A_\nu,A_\rho]\right) \right\}.
\label{action}
\eeq
It is convenient to adopt Gaussian normal coordinates for $M$, for which
the metric takes the form
\beq
ds^2 = dt^2 - h_{ij}dx^idx^j.
\label{metric}
\eeq
(A more general metric adds to the computation without changing the final
results.)  We shall further choose a set of local complex coordinates
for $\Sigma$,
\beq
h_{ij}dx^idx^j = 2h_{z\bar z}dzd{\bar z},
\eeq
in part to simplify an eventual comparison with standard Chern-Simons
results.  In a $(2+1)$-dimensional splitting, the action then becomes
\beq
S = \int\! dt \int_\Sigma\! d^2x\sqrt{h}\,\,\Tr\left\{
  -{1\over 4\gamma}F_{ij}F^{ij} - {1\over 2\gamma}F_{i0}F^{i0}
      -{k\over 4\pi}\epsilon^{ij}A_i\partial_0A_j
      +{k\over 4\pi}\epsilon^{ij}A_0F_{ij} \right\},
\label{action2}
\eeq
where
\beq
F_{ij} = \partial_iA_j - \partial_jA_i - [A_i,A_j],\quad
F_{i0} = D_iA_0 - \partial_0A_i.
\eeq
(Here, $D_i$ is the gauge-covariant derivative and $\epsilon^{ij} =
\epsilon^{0ij}/\sqrt{h}$.)

For an abelian theory, the degeneracy of energy eigenstates is already
apparent at this stage.  In the Hodge decomposition of the gauge
potential,
\beq
A = A_idx^i = { {\bf a}} + d\phi + *d\varphi,\qquad d{ {\bf a}} =
d\!*\!{ {\bf a}} = 0,
\label{Hodge}
\eeq
it is easy to check that the harmonic component ${\bf {a}}$
decouples from $\phi$ and $\varphi$.  In fact, the dynamics of the
harmonic field ${\bf a}$ is described by an effective quantum
mechanical action  equivalent to the Landau action for a charged
particle in a constant  magnetic field \cite{Wen,Kogan2}, and the
degeneracy of the Landau  states gives rise to a corresponding
degeneracy of states in the full  theory. For $G=U(1)$, the space of
gauge orbits is
\beq
\CM^{U(1)}_\Sigma=\IZ\times
\overbrace{S^1\times\cdots\times S^1}^{2g\rm\;times}
\times P(H),
\eeq
where $g$ is the genus of the surface $\Sigma$ and $P(H)$ is the
projective space associated with the transverse field $\varphi$. The
$\IZ$ connected components of $\CM^{U(1)}_\Sigma$ are parametrized
by the magnetic monopole charge of the magnetic field, and the
$(S^1)^{2g}$ describe the harmonic modes \cite{AtiyBott}.

For a nonabelian theory, this argument fails: the existence of a
Gribov problem prevents a natural generalization of the Hodge
decomposition, and the structure of the space of gauge orbits is
more complex. In particular, there is no splitting between the
topological modes (flat connections) and the propagating (transverse)
modes \cite{AtiyBott}; indeed, the space of flat connections is not
even a linear subspace of the space of nonabelian gauge fields.
On the other hand, we know that a nonabelian topologically
massive gauge  theory should reproduce the corresponding
Chern-Simons theory in the  $\gamma\rightarrow\infty$ limit
\cite{Giav}, and various other arguments  suggest a role for
Chern-Simons wave functions even when $\gamma$ is finite
\cite{Liou,Jackiw}.

To understand this situation more clearly, we shall examine topologically
massive gauge theory in the ``functional Schr\"odinger picture'' (see,
for instance, \cite{Jackiw}), in which wave functions are functionals
$\Psi=\Psi[A_i]$ and
\beq
\Pi^z = -i{\delta\ \over\delta A_z},\qquad
  \Pi^{\bar z} = -i{\delta\ \over\delta A_{\bar z}}.
\eeq
We begin by reexpressing the action \rref{action2} in canonical form.
The momenta conjugate to $A_i$ are
\beq
\Pi^i = {1\over\gamma}F^{i0} + {k\over 4\pi}\epsilon^{ij}A_j,
\eeq
while the momentum conjugate to $A_0$ vanishes identically, giving us
the nonabelian Gauss law constraint
\beq
{1\over\gamma}D_iF^{i0} + {k\over 4\pi}\epsilon^{ij}F_{ij} = 0.
\label{gl}
\eeq
This constraint is simple enough that no elaborate technology is needed
to understand the canonical theory.  In particular, the Hamiltonian
density
is
\begin{eqnarray}
H &=& \Tr\left\{\Pi^i\partial_0A_i\right\} - L \nonumber\\
 &=& -\Tr A_0\left\{ D_i\left(\Pi^i - {k\over 4\pi}\epsilon^{ij}A_j\right)
      + {k\over 4\pi}\epsilon^{ij}F_{ij}\right\} \\
 &\phantom{=}& + {1\over8\gamma}\Tr\left\{\epsilon^{ij}F_{ij}\right\}^2
      +{\gamma\over2}
         \Tr\left\{h_{ij}\left(\Pi^i- {k\over 4\pi}\epsilon^{ik}A_k\right)
         \left(\Pi^j- {k\over 4\pi}\epsilon^{jl}A_l\right)\right\},
 \nonumber
\label{Ham}
\end{eqnarray}
up to total derivatives that will vanish when we integrate to obtain the
Hamiltonian.  The first term in $H$ is simply the Gauss law constraint,
and vanishes for physical wave functions; the remainder is of the form
$B^2+E^2$.

Let us first investigate the constraint, the generator of gauge
transformations of Schr\"odinger picture states, which now takes the form
\beq
\left\{
  D_{\bar z}\left[\left(-i{\delta\ \over\delta A_{\bar z}}\right)
     + {k\over 4\pi}\epsilon^{z\bar z}A_z\right]
+ D_z\left[\left(-i{\delta\ \over\delta A_z}\right)
     - {k\over 4\pi}\epsilon^{z\bar z}A_{\bar z}\right]
+ {k\over 2\pi}\epsilon^{z\bar z}F_{z\bar z} \right\} \Psi[A_z,A_{\bar z}]
= 0.
\label{constraint}
\eeq
This expression should be contrasted with the constraint for a pure
Yang-Mills theory,
\beq
\left\{ D_{\bar z}{\delta\ \over\delta A_{\bar z}}
+ D_z{\delta\ \over\delta A_z} \right\} \Phi[A_z,A_{\bar z}] = 0,
\label{YMconstraint}
\eeq
which has as its solution any gauge-invariant functional
$\Phi[A_z,A_{\bar z}]$.  The extra terms in \rref{constraint} imply
that states in the topologically massive theory are not exactly
gauge-invariant \cite {asmitt}, transforming instead with a
one-cocycle \cite{Jackiw2}. But we can separate out the noninvariant
part of the wave function ---  essentially integrating the cocycle
condition \cite{EMSS} --- by writing \beq
\Psi[A_z,A_{\bar z}] = \exp\left\{ -{ik\over 4\pi} \int d^2x\sqrt{h}
   \epsilon^{z\bar z}A_zA_{\bar z}\right\} \chi[A_z]\Phi[A_z,A_{\bar z}],
\label{decomp}
\eeq
where the ``nontopological'' component $\Phi[A_z,A_{\bar z}]$ satisfies
the standard constraint \rref{YMconstraint} and the ``topological''
factor $\chi[A_z]$ is an arbitrary solution of the equation
\beq
\left\{ D_z{\delta\ \over\delta A_z}
  - {ik\over 2\pi}\epsilon^{z\bar z}\partial_{\bar z}A_z \right\}
  \chi[A_z] = 0.
\label{chiconstraint}
\eeq
This last expression may be recognized as the functional Schr\"odinger
equation for a Chern-Simons wave function \cite{EMSS,Bos,Dunne2}.  Its
solutions are well-understood: $\chi[A_z]$ can be described as a partition
function for a suitably gauged chiral Wess-Zumino-Witten model on
$\Sigma$ with gauge group $G$ \cite{gawkup}, and the independent
solutions are in  one-to-one correspondence with the conformal
blocks of this model. It should be stressed that the dependence of
$\chi$ on $A_z$ does not represent a breaking of diffeomorphism
invariance: a complex structure on a surface $\Sigma$, and thus a
definition of $A_z$ and $A_{\bar z}$,  is invariantly determined by
the metric $h_{ij}$.

The appearance of Chern-Simons wave functions in the solution of the
constraint is a first indication of the structure of our Hilbert
space.  The crucial question, however, is whether the
factorization \rref{decomp} is respected by the Hamiltonian.
To explore this question, observe that the
first term in $H$ is proportional to  the constraint, and vanishes on
physical wave functions.  The second  term is simply multiplicative.
The third term, on the other hand, is now
\begin{eqnarray}
{\gamma\over 2}&\Tr&
            \Biggl\{ h_{ij}\left(\Pi^i\vphantom{{k\over 4\pi}}\right.
            -\left.{k\over 4\pi}\epsilon^{ik}A_k\right)
            \left(\Pi^j- {k\over 4\pi}\epsilon^{jl}A_l\right)\Biggr\}
    \Psi[A_z,A_{\bar z}] \\
&=&\! -\gamma h_{z\bar z} \exp\left\{ -{ik\over 4\pi} \int d^2x\sqrt{h}
  \epsilon^{z\bar z}A_zA_{\bar z}\right\}\chi[A_z]
  \Tr\left\{ \left(
  {\delta\ \over\delta A_z} - {ik\over 2\pi}\epsilon^{z\bar z}A_{\bar z}
  +\J \right) {\delta\ \over\delta A_{\bar z}} \right\}
  \Phi[A_z,A_{\bar z}] \nonumber
\label{Ham2}
\end{eqnarray}
where
\beq
\J = \chi^{-1}{\delta\chi\over\delta A_z}.
\label{cb}
\eeq
We can interpret $\J$ as the expectation value of the Ka{\v c}-Moody
current of the associated gauged WZW model; as $\chi$ varies over the
Hilbert space of the Chern-Simons theory, $\J$ will vary over the
corresponding space of current blocks.

In general, the Hamiltonian thus couples the ``nontopological'' wave
function $\Phi[A_z,A_{\bar z}]$ to $\chi[A_z]$.  This mixing will be
absent only if the term
$$\J {\delta\Phi\over\delta A_{\bar z}}$$
is independent of the choice of the Chern-Simons state $\chi$.  Now,
from \rref{chiconstraint} we have
\beq
D_z \J =   {ik\over 2\pi}\epsilon^{z\bar z}\partial_{\bar z}A_z,
\eeq
which determines $\J$ up to a factor proportional to the zero-modes of
$D_z$.  Letting $\phi^i_z$ be a complete set of these zero-modes, we
can thus write
\beq
\J = \hat J^z + \sum_i
 \left( \int\! d^2w\,\sqrt{h}\, \Tr\left\{
 \langle J^w\rangle_\chi \phi^i_w\right\} \right)
 h^{z\bar z}{\bar\phi}^i_{\bar z}
\label{J}
\eeq
where $\hat J^z$ is independent of $\chi$.  Equations \rref{Ham},
\rref{Ham2} and \rref{J} then tell us that
\beq
H\Psi[A_z,A_{\bar z}] = \left(\exp\left\{ -{ik\over 4\pi}
  \int\! d^2x\,\sqrt{h}\epsilon^{z\bar z}A_zA_{\bar z}\right\}
  \chi[A_z] \right) H_0 \Phi[A_z,A_{\bar z}]
\label{HO}
\eeq
with
\begin{eqnarray}
H_0 &=& \Tr \left\{ -\gamma h_{z\bar z} \left(
 {\delta\ \over\delta A_z} - {ik\over 2\pi}\epsilon^{z\bar z}A_{\bar z}
 + \hat J^z \right) {\delta\ \over\delta A_{\bar z}}
 + {1\over2\gamma} \left( \epsilon^{z\bar z} F_{z\bar z} \right)^2
 \right\}\nonumber\\ && - \gamma \sum_i
 \left( \int\!d^2w\,\sqrt{h}\,\Tr \left\{\langle J^w\rangle_\chi \phi^i_w
 \right\}\right)\Tr{\bar\phi}^i_{\bar z}{\delta\ \over\delta A_{\bar z}},
\end{eqnarray}
and the coupling of $\Phi$ and $\chi$ is now isolated in the last term.

The presence of a Chern-Simons wave function in our decomposition will
thus give rise to degenerate energy eigenstates only in the subspace of
states for which
\beq
P^i\Phi = \int d^2z\sqrt{h}\,\Tr \left\{
 {\bar\phi}^i_{\bar z}{\delta\ \over\delta A_{\bar z}}\right\}\Phi
 = 0,\qquad [H_0,P^i]\Phi = 0,
\eeq
where the second condition is necessary to ensure that the Hamiltonian
leaves us within the appropriate space of states.  The operator $P^i$
can be interpreted as the generator of the transformation
\beq
A_z \rightarrow A_z,\qquad
 A_{\bar z} \rightarrow A_{\bar z} + {\bar\phi}^i_{\bar z},
\eeq
where the single condition on ${\bar\phi}^i_{\bar z}$ is that
$D_z{\bar\phi}^i_{\bar z} = 0$.  Hence $\Phi[A_z,A_{\bar z}]$ will be
annihilated by $P^i$ only if $A_{\bar z}$ appears solely in the
combination $\partial_z A_{\bar z} + [A_z,A_{\bar z}] = F_{z\bar z}
+ \partial_{\bar z}A_z$; that is, we must
require that
\beq
\Phi = \Phi[A_z,F_{z\bar z}].
\label{phicondition}
\eeq

If equation \rref{phicondition} is satisfied, we can set
\beq
{\delta\Phi\over\delta A_{\bar z}}
= -D_z{\delta\Phi\over\delta F_{z\bar z}},
\eeq
and by a simple calculation, the Hamiltonian density becomes
\beq
H_0 = \Tr \left\{ -\gamma h_{z\bar z} \left(
 D_z{\delta\ \over\delta A_z} - {ik\over 2\pi}\epsilon^{z\bar z}F_{z\bar z}
 \right){\delta\ \over\delta F_{z\bar z}}
 + {1\over2\gamma} \left( \epsilon^{z\bar z} F_{z\bar z} \right)^2
 \right\}.
\eeq
Using the constraint \rref{YMconstraint}, we can write this expression
as
\beq
H_0 = \Tr \left\{ -\gamma h_{z\bar z}\left(
 D_{\bar z}D_z{\delta\ \over\delta F_{z\bar z}}
 - {ik\over 2\pi}\epsilon^{z\bar z}F_{z\bar z}\right)
 {\delta\ \over\delta F_{z\bar z}}
 + {1\over2\gamma} \left( \epsilon^{z\bar z}F_{z\bar z}\right)^2
 \right\}.
\label{newH}
\eeq
It remains for us to check the commutator $[H_0,P^i]$.  Since
$P^iF_{z\bar z} = 0$, the only contribution to this commutator comes
from the explicit dependence of the covariant derivative $D_{\bar z}$
in \rref{newH} on $A_{\bar z}$, and a simple computation gives
\beq
\left[H_0,P^i\right] = -\gamma h_{z\bar z}c_{abc}{\bar\phi}^{ai}_{\bar z}
 {\delta\ \over\delta F^b_{z\bar z}}D_z{\delta\ \over\delta F^c_{z\bar z}}
\eeq
where $c_{abc}$ are the structure constants of the group $G$.  For
a generic gauge potential, this expression does not vanish, and $H_0$
takes us out of the space of states annihilated by $P^i$.  This means
that $H_0$ typically has no eigenstates that satisfy $P^i\Phi = 0$, and
the total Hamiltonian thus couples the topological and propagating modes.

If $G$ is abelian, of course, the situation is different: the
structure constants vanish, and $[H_0,P^i] = 0$.  In that case, we
can write
\beq
\Psi_{\vec n}[A_z,A_{\bar z}] = \exp\left\{ -{ik\over 4\pi}
 \int\! d^2x\,\sqrt{h}\epsilon^{z\bar z}A_zA_{\bar z}\right\}
 \chi_{\vec n}[A_z]\Phi[B], \qquad \vec n\in (\IZ_k)^g,
\label{ab1}
\eeq
with
\beq
\Phi[B]= \left\{ \begin{array}{ll}\displaystyle
\exp\left\{-{k\over 4\pi}\int\! d^2x\,\sqrt{h}B\Delta^{-1}B\right\}\xi[B]
\qquad  & \mbox{if $\displaystyle\int d^2x\sqrt{h} B = 0$}\\
0 & \mbox{if $\displaystyle\int d^2x\sqrt{h} B \ne 0$}\end{array}\right.,
\label{ab2}
\eeq
where $B = -\epsilon^{z\bar z}F_{z\bar z}$ is the magnetic field
and $\Delta = 2h^{z\bar z}\partial_{\bar z}\partial_z$.  Note that
$\Phi[B]$ vanishes for magnetic fields with non-trivial magnetic charge;
this is the canonical counterpart of the arguments of reference
\cite{Harvey}.  It is now easy to show that
\beq H_0\Phi[B]=
{1\over2}\exp\left\{-{k\over 4\pi}\int\!d^2x\,\sqrt{h}
  B\Delta^{-1}B\right\}
  \left[ \gamma{\delta\ \over\delta B}\Delta{\delta\ \over\delta B}
         + {1\over\gamma} B\Delta^{-1}
         \left(\Delta - m^2\right)B  \right]\xi[B],
\label{Ham3}
\eeq
where $m=k\gamma/2\pi$ is the usual topological mass of the photon.
We thus obtain the standard Hamiltonian for a free field
$(\gamma\Delta)^{-1/2}B$ of mass $m$; in particular, the ground state
on the plane,
\beq
\xi[B] = \exp\left\{ -{1\over2}\int\! d^2x\,d^2y\,B(x)G(x,y)B(y)
\right\}, \quad  G(x,y) = \int\!{d^2 p\over
(2\pi)^2} {\em e}^{ip(x-y)}{\sqrt{p^2+m^2}\over\gamma p^2}
\eeq
is precisely Jackiw's ground state wave function \cite{Jackiw2} for a
massive photon.

For this abelian theory, moreover, the Chern-Simons states
$\chi_{\vec n}[A_z]$ can be written explicitly \cite{BosNair}.  Let
us decompose $A_z$ as in \rref{Hodge}, and set ${\bf a}(z)=i\pi
{\bf\tilde a}\cdot({\rm Im}\,\Omega)^{-1}\cdot\omega(z)$, where $\Omega$
is the period matrix of $\Sigma$, the coefficients $\bf\tilde a$ are
constant, and the $\omega(z)$ are a basis of holomorphic differentials.
Then
\beq
\chi_{\vec n}[A_z]
 = \exp\left\{ {ik\over 4\pi} \int\!d^2z\,\sqrt{h}\left( \epsilon^{z\bar z}
   \partial_z(\phi-i\varphi)\partial_{\bar z}(\phi-i\varphi)\right) +
   {k\pi\over 2}
   {\bf\tilde a}\cdot({\rm Im}\,\Omega)^{-1}\cdot{\bf\tilde a}\right\}
   \vartheta\left[ \begin{array}{c} {\vec n}/k\\ 0 \end{array}\right]
   (k{\bf\tilde a}|k\Omega).
\label{ab3}
\eeq
It is now straightforward to combine the exponential factors from
equations \rref{ab1}, \rref{ab2}, and \rref{ab3}; the $\bf a$-independent
part is
$$ \exp\left\{ -{ik\over 4\pi}\int\!d^2x\,\sqrt{h}\phi B\right\},$$
again in agreement with Jackiw's expression for the planar case
\cite{Jackiw2}.  As anticipated in the discussion following equation
\rref{Hodge}, the quantum states associated with the harmonic modes
${\bf a}_r$, represented by $\vartheta$-functions, decouple from the
propagating modes described by the functional $\Phi[B]$. An abelian
topologically massive gauge theory is thus degenerate at all energy
levels, with a degeneracy given by the number of quantum states of
the pure Chern-Simons theory.  The degeneracy of the vacuum is the same
for all values of the coupling $\gamma$, although it is only in the
limit $\gamma\ra\infty$ that the term $\Phi[B]$ disappears and the
vacuum states become pure holomorphic Chern-Simons states.

Returning to the nonabelian case, on the other hand, we now see
that the ans\"atz of factorization of the wave function
into a holomorphic Chern-Simons part and a gauge-invariant
one does not yield a complete splitting of the Hamiltonian, and
is not useful in analyzing the spectrum of the theory.  However, the
failure of the Hamiltonian to split with such an ans\"atz is not in
itself sufficient to show a lack of degeneracy at finite $\gamma$.
In order to analyze this issue in more detail, we shall consider the
following simple perturbative argument.

As in equation \rref{decomp}, we begin by separating out a factor of
$$
\exp\left\{ -{ik\over 4\pi} \int d^2x\sqrt{h}
\epsilon^{z\bar z}A_zA_{\bar z}\right\}
$$
from the wave function (although we no longer explicitly include a
Chern-Simons wave function $\chi[A_z]$).  The Hamiltonian density is
then
\beq
H = \Tr \left\{ -\gamma h_{z\bar z} \left(
 {\delta\ \over\delta A_z} - {ik\over 2\pi}\epsilon^{z\bar z}A_{\bar z}
 \right) {\delta\ \over\delta A_{\bar z}}
 + {1\over2\gamma} \left( \epsilon^{z\bar z} F_{z\bar z} \right)^2
 \right\}.
\eeq
Observe that the two terms of the Hamiltonian have different behaviors
in the topological limit $\gamma\ra\infty$. Indeed, if we denote
\beq
H_\infty = - \Tr \left\{ \gamma h_{z\bar z} \left(
 {\delta\ \over\delta A_z} - {ik\over 2\pi}\epsilon^{z\bar z}A_{\bar z}
 \right) {\delta\ \over\delta A_{\bar z}} \right\}
 \eeq
and
\beq
V={1\over 2}\hbox{Tr}\left( \epsilon^{z\bar z} F_{z\bar z} \right)^2,
\eeq
the Hamiltonian density is $H=\gamma H_\infty + \gamma^{-1} V$;
in the topological limit, $\gamma H_\infty $ is the leading operator,
while $\gamma^{-1} V$ becomes irrelevant of order $O({1\over \gamma})$.
Note that both terms are positive, but they do not commute, so it is
not possible to simultaneously diagonalize the two operators.

The energy of any stationary state  $\psi[A_z,A_{\bar{z}}]$ will grow
as $\gamma$ in the large $\gamma$ regime unless it is annihilated by
$H_\infty$,
\beq
H_\infty\psi[A_z,A_{\bar{z}}]=0.
\eeq
Therefore as the system approaches the topological regime, its physical
quantum states are reduced to the null eigenfunctions of $H_\infty$
that satisfy the Gauss law, i.e., holomorphic functionals $\chi[A_z]$
satisfying the Gauss law constraint \rref{chiconstraint}.  We thus
recover the Chern-Simons states and the degeneracy of the Chern-Simons
theory in the topological limit $\gamma\ra\infty$.

For finite values of $\gamma$, on the other hand, since $V$ does not
commute with $H_\infty$, the holomorphic form of the eigenfunctions
is not preserved, and the degeneracy may be removed \cite{kar}.
This can be seen at leading order in perturbation theory.
The leading correction to the ground state energy generated by
the Yang-Mills
potential term $\gamma^{-1}V$ is given by the eigenvalues of
the matrix
\beq
M_{mn} = {1\over \gamma}\langle \chi_m| V[A]|\chi_n\rangle,
\eeq
where $\chi_m$, $\chi_n$ denote arbitrary states of an orthonormal
basis of the Chern-Simons theory.  As usual in field theory, the
(finite-dimensional) matrix $M$ is divergent, indicating that the
perturbative corrections to the vacuum energy require a renormalization.
The ultraviolet divergences can be partially regularized by introducing
a point splitting operator $K_\epsilon(z,{\bar{z}};w,{\bar{w}})$ between
the two curvature terms of the potential interaction,
\beq
V_\epsilon[A_z,A_{\bar{z}}]={1\over 2}\int\! d^2z\, \sqrt{h}\int\! d^2w\,
\sqrt{h}\, \Tr\left\{
\epsilon^{z\bar z} \left(D_zA_{\bar{z}} - \partial_{\bar{z}}A_z\right)
K_\epsilon(z,{\bar{z}};w,{\bar{w}})
\epsilon^{w\bar w}\left( D_w A_{\bar{w}} - \partial_{\bar{w}}A_w\right)
\right\}.
\eeq
The regulating operator $K_\epsilon$ may be chosen in a gauge-invariant
way --- for instance, it can be given by parallel transport along a
geodesic of length $\epsilon$ connecting the points $z$ and $w$,
or as the inverse of an elliptic gauge-invariant differential operator
such as $ {\cal K}_\epsilon=(I + \epsilon D_z D_{\bar{z}})^{-3}$.
For simplicity, however, we shall use the noninvariant regulator
$K_\epsilon=(I+ \epsilon \partial_z\partial_{\bar{z}})^{-3}$; gauge
invariance will be recovered after renormalization when the regulator
is removed ($\epsilon\ra 0$).

The matrix element $M_{mn}$ is then given by
\beq
M_{mn} = {1\over \gamma}\int [\delta A] \exp\left\{ -{ik\over 2\pi}
   \int\! d^2u\,\sqrt{h}\,\epsilon^{u\bar u}A_uA_{\bar u}\right\}
{\bar\chi}_m[A_{\bar{z}}]V_\epsilon[A_z,A_{\bar{z}}]\chi_n[A_z].
\label{M1}
\eeq
We can eliminate the dependence of $V_\epsilon$ on $A_{\bar{z}}$ by
functional integration by parts, using the exponential prefactor in
\rref{M1} to replace $A_{\bar z}$ by a functional derivative with
respect to $A_z$:
\begin{eqnarray}
\nonumber
M_{mn}&=&
-{1\over 2\gamma}\int [\delta A ]\exp\left\{ -{ik\over 2\pi}
   \int\! d^2u\, \sqrt{h}\,\epsilon^{u\bar u}A_uA_{\bar u}\right\}
{\bar\chi}_m[A_{\bar{z}}]\int\! d^2z\, \sqrt{h}\! \int\! d^2w\, \sqrt{h}
K_\epsilon(z,{\bar{z}};w,{\bar{w}} ) \nonumber\\
&&
\Biggl[ \Tr \left\{ \epsilon^{z\bar z}
\left(D_z A_{\bar{z}} - \partial_{\bar{z}}A_z\right)
\left({2\pi i\over k} D_w{\delta\ \over\delta A_w} +
      \epsilon^{w\bar w}\partial_{\bar{w}}A_w\right)
      - {2\pi ic_v\over k}\epsilon^{w\bar z} A_w \delta^2(z-w)A_{\bar z}
\right\} \nonumber\\
&&
 - {2\pi i\over k} \dim G\,\epsilon^{w\bar z}\partial_{\bar{z}}\partial_w
\delta^2(z-w)\Biggr]\chi_n [A_z],
\label{dlsj}
\end{eqnarray}
where $c_v$ denotes the dual Coxeter number of $G$ ($c_v=N$ for
$SU(N)$).  The first term vanishes because the Chern-Simons state
$\chi_n$ satisfies the Gauss law, but the remainder still depends on
$A_z$,$A_{\bar z}$, and does not vanish in general.

Repeating the integration by
parts to eliminate the remaining dependence on $A_{\bar z}$, we find
\begin{eqnarray}
M_{mn}&\ &=
{c_v\over 2\gamma}({2\pi \over k})^2\int [\delta A]
  \exp\left\{ -{ik\over 2\pi} \int\! d^2u\,\sqrt{h}\,
  \epsilon^{u\bar u}A_uA_{\bar u}\right\} {\bar\chi}_m[A_{\bar{z}}]
\int\! d^2z\,\sqrt{h}\!\int\!d^2w\,\sqrt{h}
  K_\epsilon(z,{\bar{z}};w,{\bar{w}})\nonumber\\
&&
\Biggl[ \Tr \left\{ A_w \delta^2(z-w){\delta\ \over\delta A_w}\right\}
 + \dim G\, (\delta^2(z-w))^2
+ {i k\dim G\over 2\pi c_v }\epsilon^{w\bar z}
\partial_{\bar{z}} \partial_w\delta^2(z-w)\Biggr]\chi_n [A_z].
\label{dddj}
\end{eqnarray}
The degeneracy of the ground state will be preserved  only if $M_{mn}$
is a diagonal matrix of the form $M_{mn}=\lambda \delta_{mn}$; if $M$
is not proportional to the identity operator, the degeneracy of the
topologically massive theory will be lower than that of the corresponding
Chern-Simons theory.

In the abelian case, equation \rref{dddj} reduces to
\begin{eqnarray}
M_{mn}&=&
{\pi i\over k\gamma}\dim G\,\int [\delta A]
  \exp\left\{ -{ik\over 2\pi} \int\!d^2u\,\sqrt{h}\,
  \epsilon^{u\bar u}A_uA_{\bar u}\right\}
{\bar\chi}_m[A_{\bar{z}}]\nonumber\\
&& \phantom{{2}}\times
\int\! d^2z\, \sqrt{h}\!\int\! d^2w\, \sqrt{h}\,\epsilon^{w\bar z}
\partial_{\bar{z}}\partial _w\delta^2(z-w)
K_\epsilon(z,{\bar{z}};w,{\bar{w}} )\chi_n [A_z]\\
&=&{\pi i\over k\gamma}\dim G\,\delta_{mn}
\int\! d^2z\,\sqrt{h}\!\int\!d^2w\,\sqrt{h}\epsilon^{w\bar z}
\partial_{\bar{z}}\partial_w\delta^2(z-w)
K_\epsilon(z,{\bar{z}};w,{\bar{w}}),
\nonumber
\label{dkkk}
\end{eqnarray}
which is proportional to the identity.  Hence the degeneracy
is preserved at least at first order in perturbation theory.  Higher
order computations will show that wave functionals become nonholomorphic
and differ from Chern-Simons wave functionals, but we know from our
previous analysis that the degeneracy is preserved at all orders of
perturbation theory.

In the nonabelian case, on the other hand, the extra term
\begin{eqnarray}
({2\pi \over k})^2{c_v\over 2 \gamma} \int&&[\delta A]\,
 \exp\left\{ -{ik\over 2\pi} \int\!d^2u\,\sqrt{h}\,
 \epsilon^{u\bar u}A_uA_{\bar u}\right\}\\
&&{\bar\chi}_m[A_{\bar{z}}]
\int\! d^2z\,\sqrt{h}\!\int\!d^2w\, \sqrt{h}\, \Tr\left\{ A_w \delta^2(z-w)
K_\epsilon(z,{\bar{z}};w,{\bar{w}} ){\delta\ \over\delta A_w}\right\}
\chi_n [A_z]\nonumber
\label{extraterm}
\end{eqnarray}
depends on $A_z$.  Hence $M_{mn}$ is not proportional to the identity,
implying that the degeneracy is broken. The extra term
\rref{extraterm} of $M_{mn}$ can be expressed in terms
of the expectation values of the Ka{\v c}-Moody currents $J^z$ and
$J^{\bar {z}}$. Because of the identity
$${\delta\ \over\delta A_z} \chi_n [A_z]= J^z \chi_n [A_z],$$
we have
\beq
M_{mn}=
 \left({2\pi \over k}\right)^3{c_v\over 2 \gamma}\int\! d^2z\,\sqrt{h}
 h_{z{\bar{z}}}  \langle {\chi}_m|J^{\bar{z}}J^z|{\chi}_n \rangle +
\lambda \delta_{mn}
\eeq
where $\lambda$ denotes the constant coefficients of the last two terms
of  expression \rref{dddj}. This connection between topologically
massive gauge theory and conformal
field theory might be exploited to explicitly evaluate the matrix
elements of $M$, thus clarifying the pattern of topological symmetry
breaking.  Although such a connection may not be useful for
higher orders in perturbation theory, its presence at the lowest order
opens up new perspectives in the application of conformal field theory
techniques to three-dimensional systems.  This topic deserves further
study.

In summary, the existence of a coupling between topological and
propagating modes in  nonabelian topologically massive gauge theories
yields an observable physical effect, the breaking of the degeneracy of
the energy levels.  This fact is not unrelated to the existence of a
shift in the effective coupling constant of the pure Chern-Simons theory
when it is viewed as the infinite mass limit of the topologically
massive theory \cite{ag, af, Giav}: in a sense,  this
shift is also due to  the coupling of the topological and propagating
modes of the massive theory.

\vspace{1ex}
\noindent
{\em Acknowledgements}: We would like to thank Ian Kogan for helpful
comments.  S.~C.\ was supported in part by the U.S.\ Department of
Energy under grant DE-FG03-91ER40674. M.~A.\ and F.~F.\ were
partially supported by  CICyT under grant AEN90-0029.


\begin{thebibliography}{99}
\bibitem{Schon} J.\ F.\ Schonfeld, \NPB{185} (1981) 157.
\bibitem{DJT} S.\ Deser, R.\ Jackiw and S.\ Templeton, \PRL{48} (1982)
  975; \Ann{140} (1982) 372.
\bibitem{Wen} X.\ G.\ Wen, \IJMPB{4} (1990) 239.
\bibitem{Wen2} X.\ G.\ Wen, \PRL{66} (1991) 802.
\bibitem{Zhang} S.\ C.\ Zhang, T.\ H.\ Hansson, and S.\ Kivelson,
  \PRL{62} (1989) 82.
\bibitem{Ho} C.-L.\ Ho, B.\ Hu, and H.\ L.\ Yu, \IJMPB{5} (1991) 1763.
\bibitem{Kogan} I.\ I.\ Kogan, \PLB{231} (1989) 377.
\bibitem{CarKog} S.\ Carlip and I.\ I.\ Kogan, \MPLA{6} (1991) 171.
\bibitem{Dunne} G.\ V.\ Dunne, R.\ Jackiw and C.\ A.\ Trugenberger,
  \PRD{41} (1990) 661.
\bibitem{Kogan2} I.\ I.\ Kogan, Comm.\ Nucl.\ Part.\ Phys. {\bf 19}
  (1990) 305.
\bibitem{Kogan3} I.\ I.\ Kogan and A.\ Yu.\ Morozov, Sov.\ Phys.\ JETP
  {\bf 61} (1985) 1.
\bibitem{kar} M.\ Asorey, in
 {\em Proceedings of the 1992 Karpacz School on Theoretical
 Physics}, {\sl J.\ Geom.\ Phys.} (to appear); Zaragoza University
 preprint DFTUZ 92.10.
\bibitem{AtiyBott} M.\ Atiyah and R.\ Bott, {\sl Philos.\ Trans.\
 Roy.\ Soc.\ London} {\bf A308} (1982) 523.
\bibitem{Giav} G.\ Giavarini, C.\ P.\ Martin, and F.\ Ruiz-Ruiz,
  \NPB{381} (1992) 222.
\bibitem{Liou} S.\ Carlip, \NPB{362}  (1991)  111.
\bibitem{Jackiw} R.\ Jackiw, in {\em Relativity, Groups and Topology II},
   edited by B.\ S.\ DeWitt and R.\ Stora (North Holland, Amsterdam,
   1984).
\bibitem{asmitt} M.\ Asorey and P.\ K.\ Mitter, \PLB{153} (1985) 147.
\bibitem{Jackiw2} R.\ Jackiw, in {\em Physics, Geometry, and Topology},
   edited by H.\ C.\ Lee (Plenum Press, New York, 1990).
\bibitem{EMSS} S.\ Elitzur, G.\ Moore, A.\ Schwimmer and N.\ Seiberg,
  \NPB{326} (1989) 108.
\bibitem{Bos} M.\ Bos and V.\ P.\ Nair, \IJMPA{5} (1990) 959.
\bibitem{Dunne2} G.\ V.\ Dunne, R.\ Jackiw and C.\ A.\ Trugenberger,
 \Ann{194} (1989) 197.
\bibitem{gawkup} G.\ Felder, K.\ Gaw\c{e}dzki, and A.\ Kupiainen, \NPB{299}
 (1988)355; \CMP{117} (1988) 127.
\bibitem{Harvey} I.\ Affleck et al., \NPB{328} (1989) 575.
\bibitem{BosNair} M.\ Bos and V.\ P.\ Nair, \PLB{223} (1989) 61.
\bibitem{ag} L.\ Alvarez-Gaum\'e, J.\ M.\ F.\ Labastida, and A.\ V.\
 Ramallo, \NPB{334} (1990) 103.
\bibitem{af} M.\ Asorey and F.\ Falceto,  \PLB{241} (1990) 31.
\end{thebibliography}
\end{document}